\documentstyle[amsfonts,aps,prl,multicol,epsf]{revtex}

\newcommand{\beq}{\begin{equation}}
\newcommand{\eeq}{\end{equation}}
\newcommand{\beqa}{\begin{eqnarray}}
\newcommand{\eeqa}{\end{eqnarray}}

\begin{document}

\title{Dynamics and delocalisation transition for an interface driven by a
uniform shear flow}

\author{Rui   D.  M.   Travasso,   Alan  J.   Bray,  Andrea   Cavagna}
\address{Department  of  Physics   \&  Astronomy,  The  University  of
Manchester, UK} \date{\today}

\maketitle

\begin{abstract}
We study the effect of a uniform shear flow on an interface separating
the  two broken-symmetry  ordered phases  of a  two-dimensional system
with  nonconserved scalar order  parameter.  The  interface, initially
flat and perpendicular to the flow, is distorted by the shear flow. We
show that there is a  critical shear rate, $\gamma_{c} \propto 1/L^2$,
(where $L$ is the system  width perpendicular to the flow) below which
the interface can sustain the shear.  In this regime the countermotion
of the interface under its  curvature balances the shear flow, and the
stretched  interface stabilizes  into a  time-independent  shape whose
form we determine  analytically. For $\gamma>\gamma_{c}$ the interface
acquires  a non-zero  velocity,  whose  profile is  shown  to reach  a
time-independent  limit  which we  determine  exactly. The  analytical
results  are checked  by  numerical integration  of  the equations  of
motion.
\end{abstract}
\begin{multicols}{2}

There is growing interest in understanding the effects of shear on the
dynamical  properties  of  statistical  systems far  from  equilibrium
\cite{onuki}.   The  most  extensively  studied case  is  perhaps  the
approach to equilibrium of a system quenched below its critical point:
domain  growth is  heavily influenced  by the  external shear  and new
dynamical  exponents  appear  \cite{shear-exp}. The  determination  of
these  exponents,  together  with  the  problem  of  the  validity  of
dynamical  scaling, are  the most  challenging tasks  in  this context
\cite{shear-th}.

The  transition  from  a  disordered,  high-temperature  phase  to  an
ordered, low-temperature  one, however, is not the  only context where
an applied  shear may  play a major  role. An  interesting alternative
problem  is  to  investigate  cases  where the  shear  may  by  itself
introduce a novel dynamics in a state otherwise ordered and stable. In
the case of spinodal decomposition  of a binary fluid, for example, we
may let the system evolve  until a stable, entirely separated state is
reached, and then  apply a shear flow normal  to the interface between
the two phases. Further evolution  will then occur, with a competition
between the shear, which tries to stretch the interface, and diffusion
of the two constituents, which tends to straighten it.
 
More generally, a  natural question in this context  is to what extent
the applied  shear is able to  perturb the stable  initial state. More
specifically, we may ask: Is there a critical value for the intensity 
of the shear, beyond which the system is unable to restore itself in a 
stable stationary state ?

In this  paper we shed some  light on the above  questions by studying
the dynamics of a flat interface subjected to a transverse shear flow.
The motion of such an interface  in the case of conserved dynamics has
been recently studied in \cite{CJV}, where the existence of a critical
value  of  the  shear  beyond  which stationarity  was  lost  was  not
reported: for the shear  rates studied, the stretched interface always
reached  a stable  steady state.   In the  present work  we  study the
deterministic dynamics of a similar interface under shear, but for the
case of nonconserved dynamics, in a system described by a scalar order
parameter. Ising-like systems, such as twisted nematic liquid crystals
\cite{Orihara},  display a  behavior  that can  be  described by  this
model.  For weak  shear we  find,  similarly to  \cite{CJV}, that  the
interface reaches a stationary  profile, in which the curvature forces
acting  on  the interface  compensate  the  shear.   In this  way  the
interface  slips relative  to the  moving boundaries,  and  acquires a
steady-state  profile with no  net velocity  in the  laboratory frame.
However, for shear rates larger than a critical value we find that the
interface cannot  sustain the strain and a  time-independent state can
no  longer  be  reached.    In  this  regime,  the  interface  departs
indefinitely  from  its  initial  condition and  becomes  delocalised.
Instead of a stationary  {\em spatial} profile, the interface acquires
a stationary {\em velocity} profile. It should be noted that the speed
of the interface at the boundaries  is always smaller than that of the
boundaries  themselves,  i.e.\  the  contact points  still  slip  with
respect to the moving boundaries, but not enough to keep the interface
stationary in the laboratory frame.

The system  we study  is a two-dimensional  strip, bounded in  the $y$
direction between  the values  $-y_{o}$ and $y_{o}$,  so its  width is
$L=2y_o$.   We consider  a uniform  shear  flow, given  by the  simple
velocity profile  ${\bf v}=\gamma y{\bf  e_x}$, where $\gamma$  is the
shear  rate and ${\bf  e_x}$ is  a unit  vector in  the $x$-direction,
which is thus the direction of  the flow. At time $t=0$, the interface
is given by the equation $x=0$,  i.e.\ it is a flat segment connecting
the boundaries and  separating two regions with opposite  value of the
order  parameter.  We  will consider  here only  the  zero temperature
dynamics of such a system.

As  a physical  boundary condition  we  assume that  the interface  is
perpendicular  to  the  boundaries  at  the  points  of  contact:  any
different condition would create an  infinite force at such points due
to the uneven curvature. We can understand better this argument in the
context of the  Ising model, by considering an  interface which is not
normal to the  boundary: the contact spin on the  acute angle side has
an excess of  neighbouring spins with opposite sign,  and it therefore
flips. This makes the interface  become normal to the interface.  This
argument  only holds  if  the microscopic  flipping  time $\tau_0$  is
smaller than the time needed by the flow to shift the spin, that is if
$\gamma  < 1/\tau_0$.  We  will find  a  critical value  of the  shear
$\gamma_c \ll  1/\tau_0$ and therefore  we can consistently  assume an
interface normal  to the boundaries at the  contact points.  Moreover,
we have checked that an Ising simulation with free boundary conditions
produces an interface which is  indeed normal to the boundaries at the
contact points.

The  deterministic  dynamics  of  a  non  conserved  order  parameter,
$\phi({\bf x},t)$, is  described by the time-dependent Ginzburg-Landau
equation  \cite{Landau},  
\beq  \tau_{o}\frac{\partial  \phi}{\partial
t}=\xi_{o}^{2}\nabla^{2}\phi -  V'(\phi) \ , 
\label{Landau}
\eeq  where $\tau_{o}$  is  the relaxation  time  for order  parameter
fluctuations  in the  bulk, $\xi_{o}$  is the  interfacial  width, and
$V(\phi)$  is a symmetric  double well  potential. Starting  from Eq.\
(\ref{Landau})  one can show  using standard  methods that  the normal
velocity of an interface separating  the two phases is proportional to
the local  curvature.  This is the Allen-Cahn  equation \cite{AC}, $v=
-D\  \nabla\cdot  {\bf  n}$,   where  $\nabla\cdot  {\bf  n}$  is  the
curvature,  $D=\xi^{2}_{o}/\tau_{o}$  is  the diffusion  constant  for
fluctuations in the bulk phases, and  ${\bf n}$ is the local normal to
the interface.  The Allen-Cahn equation is actually  more general than
the  specific  dynamical  equation  (\ref{Landau}),  since  it  simply
expresses the dynamics of an interface driven by its surface tension.

In the absence of shear, a flat interface is stable. When a shear is 
applied, however, the Allen-Cahn equation has to be modified: 
\beq
v=-D\ \nabla\cdot {\bf n} + \gamma y\,{\bf e_x}\cdot{\bf n} \ ,
\label{allen}
\eeq  where  the  additional  term  represents the  advection  of  the
interface by the flow, leading to a distortion of its initially planar
form.   Describing the  interface  profile by  the function  $x(y,t)$,
which gives  the displacement in the  flow direction as  a function of
the position  in the direction transverse  to the flow, it  is easy to
show that the  divergence of the normal is  given by: \beq \nabla\cdot
{\bf n}= -\frac{\partial_{y}^{2}x}{[1+(\partial_{y}x)^{2}]^{3/2}},
\label{divn}
\eeq
while the velocity, $v$, of the interface can be written as
\beq
v=\frac{\partial_{t} x}{[1+(\partial_{y}x)^{2}]^{1/2}} \ , 
\label{vx}
\eeq  where $\partial_t x$  is the  velocity of  the interface  in the
direction parallel to  the flow. Combining (\ref{allen}), (\ref{divn})
and  (\ref{vx})  we  obtain  an  equation for  the  interface  profile
$x(y,t)$:                                                          \beq
\partial_{t}x=D\,\frac{\partial_{y}^{2}x}{1+(\partial_{y}x)^{2}}      +
\gamma y \ .
\label{eq1'}
\eeq Prior to further analysis,  it is convenient to rescale space and
time, \beq X\equiv x/y_{o} \quad , \quad Y\equiv y/y_{o} \quad , \quad
\tau\equiv   t\,   D/  y_o^2   \   ,   \nonumber\\   \eeq  where   now
$Y\in[-1,1]$. Note that in this way we are also introducing a rescaled
interface velocity $V=v\,y_o/D$.  In  terms of the rescaled variables,
Eq.\               (\ref{eq1'})               reads               \beq
\partial_{\tau}X=\frac{\partial_{Y}^{2}X}{1+(\partial_{Y}X)^{2}}      +
\alpha Y \quad , \quad \alpha\equiv \gamma y_o^2 /D \ .
\label{eq1}
\eeq Due to the rescaling,  all the dimensionfull parameters have been
absorbed into the dimensionless  effective shear rate $\alpha$.  Given
the geometry  of the problem, we  clearly expect $X(Y,\tau)$  to be an
odd  function of  $Y$. We  will therefore  limit our  analysis  to the
domain $Y\in[0,1]$.

In  order   to  find  a  stationary  solution   $X_s(Y)$  of  equation
(\ref{eq1}),  we  set $\partial_{\tau}X=0$,  to  obtain (where  primes
indicate     derivatives     with     respect     to     $Y$)     \beq
\frac{X_{s}''}{1+(X_{s}')^{2}}=-\alpha Y\ ,
\label{indy}
\eeq with  boundary conditions \beq X'_s(1)  = 0 =  X_s(0), \eeq where
the first condition follows from the requirement that the interface be
perpendicular to  the boundaries,  and the second  from the  fact that
$X(Y)$  is  an  odd  function.   Integrating once,  and  imposing  the
boundary    condition   at    $Y=1$,    gives   \beq    X_{s}'(Y)=\tan
\left[\frac{\alpha}{2}(1-Y^{2})\right] \ .
\label{eq4}
\eeq
A second integration, incorporating the boundary condition at $Y=0$, 
gives the stationary interface profile,
\beq
X(Y)=\int_{0}^{Y}\tan
\left[\frac{\alpha}{2}(1-z^{2})\right] {\mathrm d}z\ . 
\label{eq3}
\eeq  This interface  profile is  plotted  in Figure  1 for  different
values of the effective shear rate.

From equation  (\ref{eq4}) we see  that the function $X_{s}(Y)$  has a
maximum in  its derivative at $Y=0$. When  $\alpha=\pi$ the derivative
diverges at  this point indicating  that the interface is  parallel to
the   system  boundaries.   A   value  $\alpha>\pi$   would  imply   a
discontinuous  derivative  $X'_s(Y)$  and  thus an  interface  profile
$X_s(Y)$ with two  cusps: such a profile would  be unphysical, because
of the infinite curvature at  the cusps. We conclude that for $\alpha>
\pi$ no stationary solution is  possible. This implies that there is a
critical value of the  dimensionless shear rate, 
\beq 
\alpha_{c}=\pi \,  
\eeq 
i.e.\ a critical  shear  rate 
\beq  
\gamma_c =  \frac{4\pi D}{L^2} \ .  
\eeq 

\begin{figure}
\narrowtext\centerline{\epsfxsize\columnwidth\epsfbox{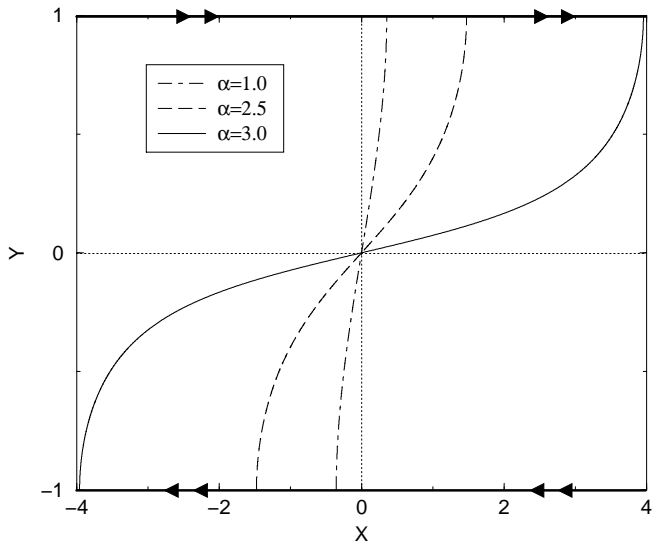}}
\caption{Stationary   interface   profile   as   given  by  equation
(\ref{eq3}), for  three different values  of the effective  shear rate
$\alpha$. The initial  configuration of the interface is  given by the
flat $X=0$ segment.}
\end{figure}

For $\alpha>\alpha_c$ (or $\gamma>\gamma_c$) we must
find a  time-dependent solution of the full  equation (\ref{eq1}).  In
terms  of the  microscopic parameters  of  the systems,  we have  \beq
\gamma_c  =  \frac{4\pi  \xi_0^2}{L^2}  \, \frac{1}{\tau_0}  \eeq  and
therefore $\gamma_c \ll 1/\tau_0$ as  long as the interfacial width is
much smaller than  the size of the system,  a condition that obviously
holds  for any  reasonable system.  Therefore, as  anticipated  in the
Introduction, our assumption of  an interface normal to the boundaries
at  the contact  points  is fully  justified  in the  totality of  the
interesting $\gamma$ regime.

For $\alpha\to\alpha_c$ the contact  point $X(1)$ goes to infinity and
the interface is  no longer localized in a finite  region of space. We
have, 
\beq X(1)  \sim \sqrt{ \frac{\alpha_c}{\alpha-\alpha_c}} \quad ,
\quad \alpha \to \alpha_c \ .
\label{contact}
\eeq In order to obtain this  result it must be noted that for $\alpha
\to  \alpha_c$  the integral  in  equation  (\ref{eq3}) is  completely
concentrated near the  origin, $z = 0$, and  it therefore depends very
weakly on  the value  of $Y$,  as long as  $Y$ is  non-vanishing. This
means that for $\alpha\to \alpha_c$ not only the contact point $X(1)$,
but {\it all} the points of the interface $X(Y)$ with non-zero $Y$ are
found at the same position given by (\ref{contact}).

Before considering the $\alpha>\alpha_c$ regime, we want to check that
the time-independent solution we have found is indeed an attractor for
the interface  dynamics, when starting from  the initial configuration
$X(Y,0)=0$.   To  do  this,   Eq.\  (\ref{eq1})  was  discretized  and
integrated  numerically.  In  Figure 2  we  plot the  position of  the
contact point,  $X_s(1)$, of the  asymptotic time-independent solution
as a  function of  the shear  rate, in order  to compare  our analytic
result,  Eq.\  (\ref{eq3}), with  the  numerical  integration of  Eq.\
(\ref{eq1}).   The agreement  is  good, and  it  can be  seen that  it
improves as the continuum limit is approached.  The conclusion is that
the stationary state given by  Eq.\ (\ref{eq3}) is indeed an attractor
for the interface dynamics.

\begin{figure}
\narrowtext\centerline{\epsfxsize\columnwidth\epsfbox{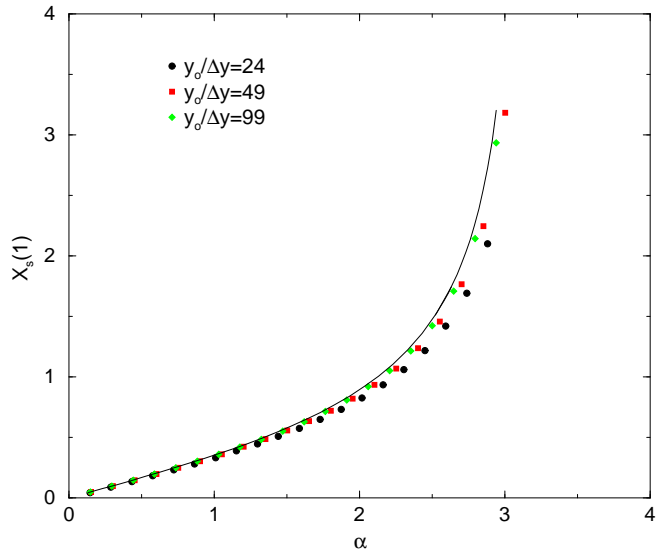}}
\caption{Numerical integration  of equation (\ref{eq1})  (symbols) and
analytical  result (full line).  The numerical  integration approaches
the analytical result as the spatial discretization scale, $\Delta y$,
decreases.}
\end{figure}

When  $\alpha  > \alpha_{c}$  the  time-independent  equation for  the
interface has no physical solution  and the interface must move with a
nonzero velocity  parallel to the boundaries.  A reasonable assumption
is that,  in the large-time limit,  the velocity of  every point along
the    interface    is    time-independent,    i.e.     $\partial_\tau
X(Y,\tau)\equiv V(Y,\tau)  \to V_\infty(Y)$.  Such  a velocity profile
for  the  interface  implies  that  the  nonlinear  term  in  equation
(\ref{eq1})  must be  time-independent in  the asymptotic  limit. This
condition must  be satisfied in  two different ways, according  to the
different regions of the $Y$-domain.

$\bullet  \  Y\sim  0:  \  \  $  From  the  time-independent  equation
(\ref{indy}) we see that $X_s''(0)=0$ for all $\alpha<\alpha_c$. It is
natural then  to assume that  even in the phase  $\alpha>\alpha_c$ the
interface is  asymptotically flat close  to the centre, such  that the
nonlinear  term  vanishes in  this  region  and  the solution  of  the
equation is trivial, \beq X(Y,\tau)=\alpha Y\tau\ .
\label{e2}
\eeq
This solution clearly satisfies Eq.\ (\ref{eq1}), and the boundary 
condition $X(0,\tau)=0$. However, it cannot be correct close to the 
boundary at $Y=1$ because it does not ratify the boundary condition 
$\partial_{Y}X(1,\tau)=0$. We will find that it is correct in a
domain $0 \le Y \le Y_1$, with $Y_1$ to be determined below. 
 
$\bullet \  Y\sim 1: \  \ $ At  $Y=1$, the interface must  satisfy the
boundary condition $\partial_{Y}X(1,\tau)=0$, and therefore we need to
keep  the nonlinear term  in the  equation of  motion. However,  as we
already noted,  this term  must become independent  of time  for large
times.  This can be  achieved by  requiring that  
\beq X(Y,\tau)\equiv
f(Y) +  V_o\tau \  ,\ \ \  Y\sim 1,  
\eeq 
so that  the portion  of the
interface close to  the system boundary moves with  a velocity $V_{o}$
independent of $Y$. This  solution also satisfies Eq.\ (\ref{eq1}). We
shall  find  that   it  is  correct  in  the  domain   $Y_1  <  Y  \le
1$.        Integrating         equation        (\ref{eq1})        with
$\partial_{\tau}X(Y,\tau)=V_{o}$  and  $\partial_{Y} X(1,\tau)=0$,  we
obtain                     
\beq  \partial_{Y}X=\tan
\left[\frac{\alpha}{2}(1-Y^{2})-V_{o}(1-Y)\right]\ .
\label{e3}
\eeq

The two solutions eqs. (\ref{e2}) and (\ref{e3}) must be matched 
at the border between the two different regions. Let us define 
region I: $0\le Y \le Y_{1}$ and region II: $Y_{1} \le Y \le 1$. 
The matching point $Y_1$ will be calculated later. 
We can write the complete interface velocity profile, for 
$\tau \to \infty$, as
\beq
V_\infty(Y)=
\left\{
\begin{array}{ll}
\alpha Y  & \quad I \\
V_{o}=\alpha Y_{1} & \quad II
\end{array}
\right.
\label{eq10}
\eeq
For the $Y$ derivative of the interface profile, for 
$\tau \to \infty$, we have
\beq
\partial_{Y}X_{\infty}(Y)=
\left\{
\begin{array}{ll}
\infty & \quad I \\
\tan\left[\frac{\alpha}{2}(1-Y^{2})-V_{o}(1-Y)\right]
& \quad II \\
\end{array}
\right.
\label{eq11}
\eeq Furthermore,  note that  the value of  $V_o$ is known,  from Eq.\
(\ref{eq10}), once  $Y_1$ is determined.  In order  to calculate $Y_1$
we  have to  match  the solutions  in  the different  regions at  this
point. Note  that the derivative, $\partial_YX$, has  to be continuous
at this point  to avoid an infinite curvature.   From (\ref{eq11}), we
see that this requires the argument  of the tangent in region II to be
equal to $\pi/2$ when $Y=Y_1$.  This gives      
\beqa      
Y_{1} &=&1-\sqrt\frac{\alpha_{c}}{\alpha}    \nonumber    \\    
V_{o}    &=& \alpha\left(1-\sqrt\frac{\alpha_{c}}{\alpha}\right) \ .
\label{v}
\eeqa 
Note  that the  velocity $V_o$ of  the interface at  the contact
point, $Y=1$,  is smaller than the  velocity of the flow  at this same
point which, in our dimensionless variables, is equal to $\alpha$. Our
analytical  solution  was  tested  by  numerical  integration  of  the
dynamical equation, and the results  are shown in Figure 3.  Numerical
and analytical results are fully consistent.

The  convergence  of the  interface  velocity  at  the contact  point,
$\partial_\tau  X(1,\tau)   \rightarrow  V_{o}$  for  $\tau\rightarrow
\infty$,  was  found to  be  very  slow.  The function  $\partial_\tau
X(1,t)$ is monotonically decreasing in  time: initially it is equal to
the flow  velocity at  the contact point  $\alpha$, but  eventually it
must decrease, in  order to develop a curvature  to keep the interface
perpendicular to the  boundary. A closer inspection of  Figure 3 shows
that for any  finite time the cusp in the  velocity profile is rounded
out and this function is smooth.

\begin{figure}
\narrowtext\centerline{\epsfxsize\columnwidth\epsfbox{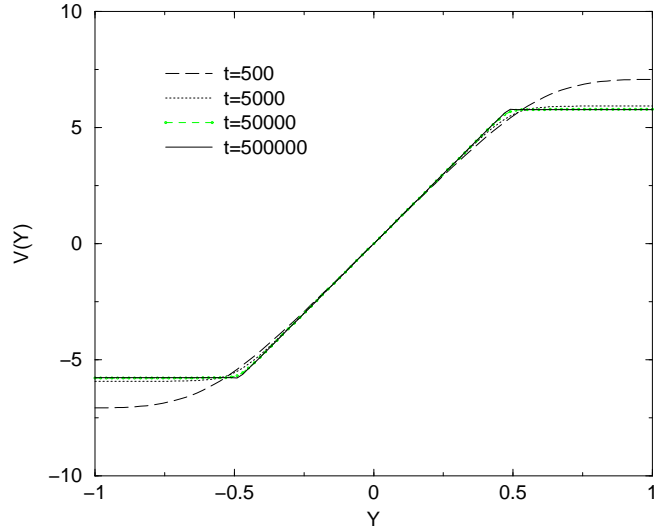}}
\caption{The velocity profile of the interface as a function of $Y$, 
with $\alpha=12$. It can be clearly seen that for large times the 
velocity profile converges towards the function $V_\infty(Y)$ of 
eq. (\ref{eq10}).}
\end{figure}

However, for very  large times, it is not  unreasonable to approximate
the actual velocity profile with  the same piecewise linear form as in
eq.(\ref{eq11})  (see Figure  4). Of  course, we  have to  introduce a
time-dependent  value,  $Y_1(\tau)$,  for  the  matching  point,  with
$Y_1(\tau)\to Y_1$ for  $\tau\to\infty$.  Within this approximation we
can therefore write 
\beq \partial_{Y}X(Y,\tau)\approx \left\{
\begin{array}{ll}
\alpha \tau  & \quad I \\
\tan\left[\frac{\alpha}{2}
(1-Y^{2})-V_{o}(\tau)(1-Y)\right] & \quad II \\
\end{array}
\right.
\label{eq12}
\eeq       
where $V_{o}(\tau)=\alpha Y_{1}(\tau)$ and
$Y_{1}(\tau)=Y_{1}+\delta Y_{1}(\tau)$. The  new value of the matching
point $Y_{1}(\tau)$ is fixed, as  usual, by imposing the continuity of
$\partial_{Y}X(Y,\tau)$. This gives    
\beq  
\tan^{-1}(\alpha \tau)\approx\frac{\pi}{2}-\frac{1}{\alpha \tau}
\approx\frac{\alpha}{2}[1-Y_{1}(\tau)]^{2}         
\eeq        
Setting $Y_{1}(\tau)=Y_{1}+\delta Y_{1}$ and recalling that
$\frac{\alpha}{2}(1-Y_{1})^{2} =\frac{\pi}{2}$  from the definition of
$Y_{1}$  in  equation (\ref{v}),  we  get  
\beq  
\delta Y_{1}  \approx
\frac{1}{\alpha^{2}(1-Y_{1})}\frac{1}{\tau}   \   ,   
\eeq   
and   for $V(1,\tau)$                    
\beq 
V(1,\tau)\approx V_{o}+
\frac{1}{\sqrt{\alpha\alpha_{c}}}\frac{1}{\tau} \ .  
\eeq

\begin{figure}
\narrowtext\centerline{\epsfxsize\columnwidth\epsfbox{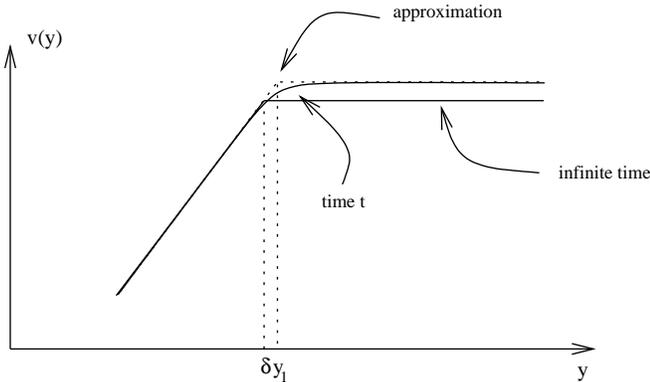}}
\caption{The nature of the approximation leading to equation 
(\ref{eq12}).}
\end{figure}

In order to check this last result numerically, it is convenient to
compute the time derivative of $V(1,\tau)$, which does not contain 
the constant $V_o$. We have
\beq
{\partial_\tau V}(1,\tau)\approx -\frac{1}{\sqrt{\alpha\alpha_{c}}}
\ \frac{1}{\tau^{2}} \ .
\label{final}
\eeq This  last result is  tested in Figure  5, where we may  see that
again numerical  integration of the dynamical  equation and analytical
results coincide.

\begin{figure}
\narrowtext\centerline{\epsfxsize\columnwidth\epsfbox{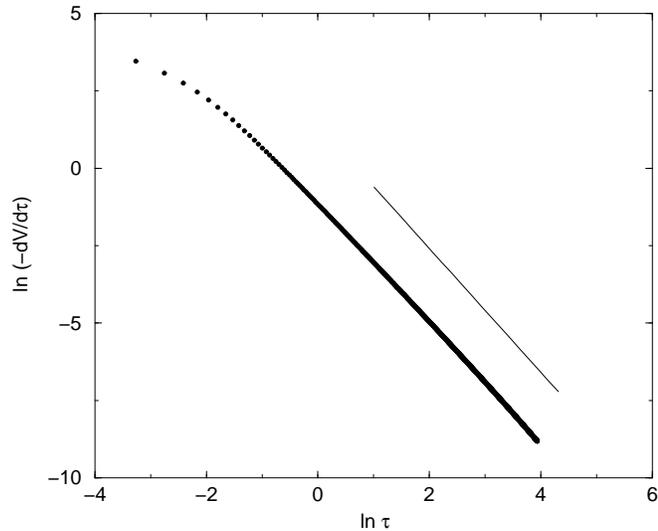}}
\caption{The  quantity  $\partial_\tau  V(1,\tau)$  as a  function  of
$\tau$  in a  log-log  plot.  The  full  line has  slope  $-2$, for  a
comparison with the analytical result eq.(\ref{final}).}
\end{figure}

In this  paper we have  analytically and numerically studied  the zero
temperature  dynamics of an  interface subject  to a  transverse shear
flow, in the case of  nonconserved dynamics. We find a critical value,
$\alpha_c=\pi$,  of the  dimensionless  shear rate,  $\alpha =  \gamma
y_o^2/D$,  beyond which  a  steady  state cannot  be  reached and  the
interface moves with  a constant velocity. In terms  of the thickness,
$L=2y_o$, of the sample, this gives $\gamma_c = 4\pi D/L^2$, i.e.\ the
transition occurs at lower shear rates for wider systems.
 
While  no such  critical value  was  reported for  the conserved  case
analyzed in  \cite{CJV}, it  is unclear  to us why  there should  be a
major difference  in this  respect between conserved  and nonconserved
dynamics. Indeed, in  the latter case, the dynamics  is slower, due to
the conservation constraint, and the  system should be less capable of
sustaining the  shear than in  the nonconserved case.   We conjecture,
therefore, that a similar transition occurs for systems with conserved
dynamics.

\acknowledgements It  is a  pleasure to thank  F. Thalmann  for useful
discussions.  This  work was supported by EPSRC  under grant GR/L97698
(AC and  AJB), and by  Funda\c c\~ao para  a Ci\^encia e  a Tecnologia
under program PRAXIS XXI, grant BD/21760/99 (RDMT).

\end{multicols}
\end{document}